\documentclass{cernrep}
\usepackage{wrapfig} 
\usepackage{texnames}
\usepackage[T1]{fontenc}
\begin{document}
\title{Practical Statistics for Particle Physicists}
 
\author{H. B. Prosper}

\institute{Florida State University, Tallahassee, USA}  

\maketitle 

\begin{abstract}
We introduce a few of the key ideas of statistical analysis using two real-world examples to illustrate how these ideas are used in practice. 
\end{abstract}

\section{Introduction}
These
lectures introduce to two broad classes of theories of inference, the frequentist
and Bayesian approaches. Two points should be made immediately. The first is that  there is no such thing as ``the" answer in statistics. Instead there are answers based on
assumptions on which reasonable people may disagree.  Second, none of the current theories of inference is perfect. 
It is worth appreciating these points in order to avoid fruitless arguments
that cannot be resolved because they are ultimately
about intellectual taste and not  mathematical correctness.

For in-depth expositions of statistical analysis,  we
highly recommend the excellent books on statistics written for physicists, by physicists~\cite{Lyons, James, Cowan, Barlow}, and the very insightful book on the history of the
ideas by Chatterjee~\cite{Chatterjee}.

\section{Lecture 1: descriptive statistics, probability and likelihood}
\subsection{Descriptive statistics}
\label{sec:statistics}
Suppose we have a sample of $N$ data $X = x_1, x_2, \cdots, x_N$. It is often useful to summarize these data with a few numbers called statistics. 
A \textbf{statistic} is any number that can be calculated  from the data and known parameters.  For example, $t = (x_1 + x_N)/2$ is a statistic, but if the value of $\theta$ is unknown in $t = (x_1 - \theta)^2$ then the latter is not  a statistic. However,  we particle physicists tend to refer to \emph{any} function of the data as a statistic including those that contain unknown parameters. 

The two most important  statistics are
\begin{align}
	\label{eq:xbar}
	&\textrm{the {\bf sample mean} (or average)} & \bar{x} 		& =  \frac{1}{N} \sum_{i=1}^N x_i, \\
		\label{eq:xvar}	
	&\textrm{and the {\bf sample variance}} & s^2 	& =  \frac{1}{N} \sum_{i=1}^N (x_i - \bar{x})^2,
	\nonumber\\
	& & & = \frac{1}{N} \sum_{i=1}^N x_i^2 - \bar{x}^2, \nonumber\\
	& & & = \overline{x^2} - \bar{x}^2.
\end{align}
The sample average is a measure of the center of the distribution of the data, while the sample variance is a measure of its spread. Statistics that merely characterize the data are
called \textbf{descriptive statistics}, of which the sample average and variance are the most
important. 

Descriptive statistics can always be calculated because they depend only on a data sample $X$. We now consider numbers that cannot be calculated from the data alone.  Imagine the repetition, infinitely many times, of the data generating system that yielded our data sample $X$, thereby creating an infinite set of data sets. We shall refer to the data generating system as an experiment and the
infinite set of the results of the experiments as an infinite ensemble. This
is clearly an 
abstraction.  

The most common operation to perform on an ensemble is to compute the \textbf{ensemble average} of the statistics, which yield numbers such as the following.
\begin{align}
& \textrm{Ensemble average} 		& & \langle x \rangle				\nonumber\\
& \textrm{Mean}				& \mu				\nonumber\\
& \textrm{Error}					& \epsilon & = x - \mu	\nonumber\\
& \textrm{Bias}					& b & = \langle x \rangle - \mu	 	\nonumber\\
& \textrm{Variance}				& V & = \langle (x - \langle x \rangle)^2 \rangle	\nonumber\\
& \textrm{Standard deviation}		& \sigma & = \sqrt{V}		\nonumber\\
& \textrm{Mean square error}		& \text{MSE} & = \langle (x - \mu)^2 \rangle	\nonumber\\
& \textrm{Root MSE}				& \textrm{RMS} & = \sqrt{\textrm{MSE}}
\label{eq:ensemble}	
\end{align}
None of these numbers can be calculated from data because the data
needed do not objectively exist. Even
in an experiment simulated on a computer, there are very few of these numbers we can
calculate. If we know the mean $\mu$, perhaps because we have chosen its value, we can certainly calculate the error $\epsilon$ for any simulated datum $x$. But, we can only \emph{approximate} the
ensemble average $\langle x \rangle$, bias $b$, variance $V$, and MSE, since the ensembles available either on our computers or in the real world are
always finite. The point is that the numbers that characterize the infinite ensemble are also abstractions.

The MSE is the most widely used
measure of the closeness of an ensemble of numbers to some parameter $\mu$. The square root of the MSE is called the root mean square (RMS)\footnote{Sometimes, the RMS and standard deviation are using interchangeably. However, the RMS is computed with respect to $\mu$, while the standard deviation is computed with respect to the ensemble average $\langle x \rangle$. The RMS and standard deviations are identical only if the bias is zero.}. The MSE can be written as
\begin{align}
	\textrm{MSE} & = V + b^2,	\\ & \framebox{\textbf{Exercise 1: } \textrm{Show this}}\nonumber
\end{align}
the sum of the variance and the square of the bias, 
 a very important result with practical consequences. For example, suppose that $\mu$ represents the mass of the Higgs boson and $x$ is a complicated function that is considered an \textbf{estimator} of the mass. An estimator is any
function, which when data are entered into it, yields an \textbf{estimate} of the quantity
of interest.

As noted, many of the numbers listed in Eq.~(\ref{eq:ensemble}) cannot be calculated because
the information needed is unknown. This is
true, in particular, of the bias. However, sometimes it is possible to relate the bias to another ensemble quantity. Consider the ensemble average of the sample variance, Eq.~(\ref{eq:xvar}),
\begin{align}
	\langle s^2 \rangle 	& = \langle \overline{x^2} \rangle - \langle \bar{x}^2 \rangle, \nonumber\\
			& = V - \frac{V}{N}, \nonumber\\
			&  \framebox{\textbf{Exercise 2a:} Show this} \nonumber		
\end{align}

\centerline{		
\framebox{\parbox{0.5\textwidth}{\textbf{Exercise 2b:} Use the method {Rndm()} of the {\tt Root}\\ class {\tt TRandom3} to approximate the quantities in Eq.~(\ref{eq:ensemble}). }}	
}

\subsection{Probability}
\label{sec:prob}
When the weather forecast specifies that there is a 80\% chance of snow tomorrow at CERN, most people
have an intuitive sense of what this means. Likewise, most people have an intuitive 
understanding of what it means to say that there is a 50-50 chance for a tossed coin to land
heads up. Probabilistic ideas are thousands of years old, but, starting in the sixteenth century   these ideas were formalized into increasingly rigorous mathematical theories of probability. 
In the theory formulated by Kolmogorov in 1933, $\Omega$ is
some fixed mathematical space, $E_1, E_2, \cdots \subset \Omega$ are subsets (called events) defined in some
reasonable way\footnote{If $E_1, E_2, \cdots$ are meaningful subsets of $\Omega$, so to is the complement $\overline{E}_1, \overline{E}_2, \cdots$  of each, as are
countable unions and intersections of these subsets.}, and $P(E_j)$ is a number 
associated with subset $E_j$. These numbers satisfy the
\begin{align*}
	\textbf{Kolmogorov Axioms}			\\
	& \quad 1.	 \quad P(E_j) \geq 0 		\\
	& \quad 2. \quad P(E_1 + E_2 + \cdots) = P(E_1) + P(E_2) + \cdots
	\quad\textrm{for disjoint subsets} \\
	& \quad 3. \quad P(\Omega) = 1.
\end{align*}
Consider two subsets $A = E_1$ and $B = E_2$. The quantity $AB$ means $A$ \emph{and} $B$, while $A + B$ means $A$ \emph{or}
$B$, with associated probabilities $P(AB)$ and $P(A+B)$, respectively. Kolmogorov assumed, not unreasonably given the intuitive origins of probability, that probabilities sum to unity; hence the axiom $P(\Omega) = 1$. However, this assumption can be dropped so that probabilities remain meaningful even if $P(\Omega) = \infty$~\cite{Taraldsen}. 

\begin{wrapfigure}{R}{0.5\textwidth}
\centering\includegraphics[width=0.5\textwidth]{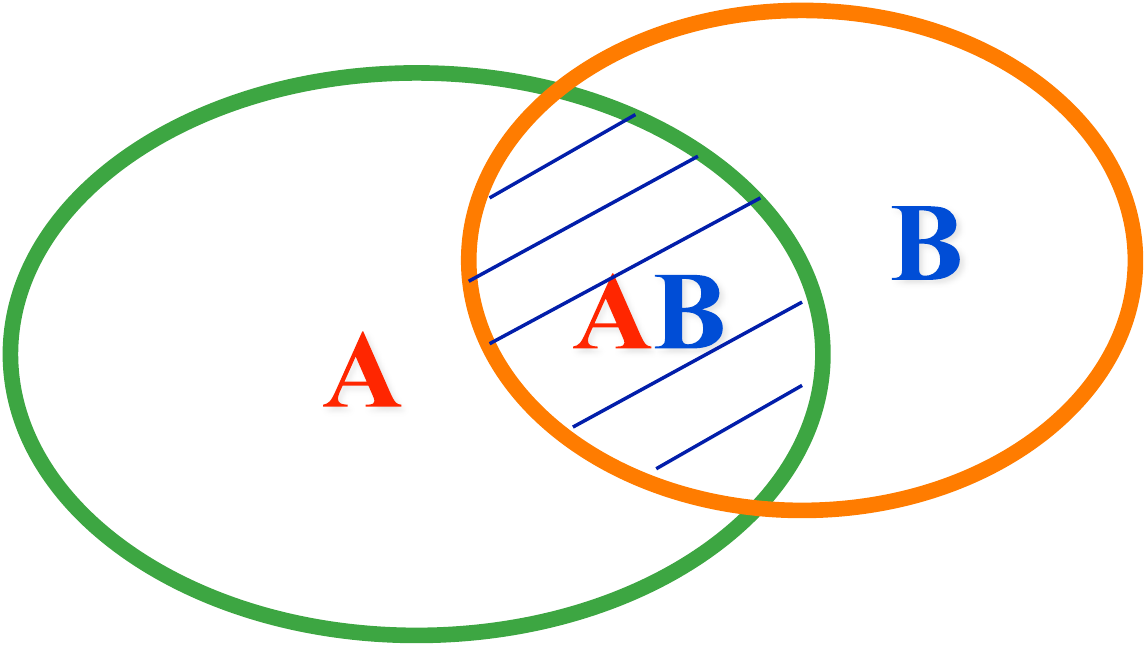}
\caption{Venn diagram of the sets $A$, $B$, and $AB$. $P(A)$ is the probability of $A$, while $P(A|B) = P(AB) / P(B)$ is the probability of $AB$ relative to that 
of $B$, i.e.,  the probability of $A$ given the condition $B$.} 
\label{fig:venn}
\end{wrapfigure}

Figure~\ref{fig:venn} suggests another probability, namely, the number $P(A|B) = P(AB) / P(B)$, called the \textbf{conditional probability} of
$A$ given $B$. This permits statements such as: ``the probability that this track was
created by an electron given the measured track parameters" or ``the probability to observe
17 events given that the mean background is 3.8 events".
Conditional probability is a very powerful idea, but the term itself is misleading. It implies that there are two kinds of
probability: conditional and unconditional. In fact, \emph{all} probabilities are conditional in
that they always depend on a specific set of conditions, namely, those that
define the space $\Omega$. It is entirely possible to embed a family of subsets of $\Omega$ 
into another space $\Omega^\prime$ which assigns to each family member a different
probability $P^\prime$. A probability is defined only relative to some space of possibilities $\Omega$.

$A$ and $B$ are said to be mutually exclusive if $P(AB) = 0$, that is, if the truth
of one denies the truth of the other. They are said to be exhaustive if $P(A) + P(B) = 1$. 
Figure~\ref{fig:venn} suggests the theorem
\begin{align}
	P(A + B) = P(A) + P(B) - P(AB), \\
	\framebox{\textbf{Exercise 3:} Prove theorem}\nonumber
\end{align}
which can be deduced from the rules given above. Another useful theorem is 
an immediate consequence of the commutativity of ``anding" 
 $P(AB) = P(BA)$ and the definition of $P(A|B)$,
namely,
\begin{align}
\textbf{Bayes Theorem} \nonumber\\
&P(B|A ) = \frac{P(A|B) P(B)}{P(A)}, 
\end{align} 
which provides a way to convert the probability $P(A|B)$ to the probability $P(B|A)$. 
Using Bayes theorem, we can, for example,
deduce the probability $P(e|x)$  that a particle is an electron, $e$, given a set of measurements, $x$,  from the
probability $P(x|e)$ of a set of measurements given that the particle is an electron.

\subsubsection{Probability distributions}
In this section, we illustrate the use of these rules to derive more complicated 
probabilities. First we start with a definition:
\begin{quote}
A \textbf{Bernoulli trial}, named after the Swiss mathematician Jacob Bernoulli (1654 -- 1705), is an experiment with only two possible outcomes: $S = \textrm{success}$ or $F = \textrm{failure}$. 
\end{quote}

\begin{quote}
\paragraph*{Example} Each collision between protons at the Large Hadron Collider (LHC) is a Bernoulli trial in which something interesting happens ($S$) or does not ($F$). Let $p$ be
the probability of a success, which is assumed to be the \emph{same for each trial}. Since $S$
and $F$ are exhaustive, the probability of a failure is $1 - p$. For  a given order $O$ of $n$
proton-proton collisions and exactly $k$ successes, and therefore exactly $n - k$ failures, the probability $P(k, O , n, p)$ is given by
\begin{align}
	P(k, O, n, p) = p^k (1 - p)^{n - k}.
\end{align}
If the order $O$ of successes and failures is judged to be irrelevant, we can eliminate the order
from the problem by summing over all possible orders,
\begin{align}
	P(k, n, p) = \sum_O P(k, O, n, p) = \sum_O p^k (1 - p)^{n - k}.
	\label{eq:Pkn}
\end{align}
This procedure is called \textbf{marginalization}. It is one of the most important operations in probability calculations. Every term in the sum in Eq.~(\ref{eq:Pkn}) is identical and there
are $\binom{n}{k}$ of them. This yields the \textbf{binomial distribution},
\begin{align}
	\textrm{Binomial(k, n, p)} \equiv \binom{n}{k} p^k (1 - p)^{n - k}.
\end{align}
By definition, the mean number of successes $a$ is given by
\begin{align}
	a 	& = \sum_{k=0}^n k \, \textrm{Binomial(k, n, p)}, \nonumber\\
		& = p n. \\
		& \framebox{\textbf{Exercise 4:} Show this} \nonumber
\end{align}
At the LHC $n$ is a number in the trillions, while for successes of interest such as the creation of a Higgs boson
the probability $p << 1$. In this case, it proves convenient to consider the 
limit $p \rightarrow 0, n \rightarrow \infty$ in such a way that $a$ remains constant. In
this limit
\begin{align}
	\textrm{Binomial(k, n, p)}	& \rightarrow e^{-a} a^k / k! , \nonumber\\
						& \equiv \textrm{Poisson}(k, a).\\
						& \framebox{\textbf{Exercise 5:} Show this} \nonumber
\end{align}
\end{quote}

\bigskip

\noindent
Below we list the most common probability distributions.
\begin{align}
&\textbf{Discrete distributions}\nonumber\\
& \textrm{Binomial}(k, n, p)		& \binom{n}{k} p^k (1 - p^{n-k}	\nonumber\\
& \textrm{Poisson}(k, a)			& a^k \exp(-a) / k!			\nonumber\\
& \textrm{Multinomial}(k, n, p)		& \frac{n!}{k_1!\cdots k_K!} \prod_{i=1}^K p_i^{k_i},		\quad \sum_{i=1}^K p_i = 1, \sum_{i=1}^K k_i = n		\nonumber\\
&\textbf{Continuous densities}\nonumber\\
& \textrm{Uniform}(x, a)			& 1 / a	\nonumber\\
& \textrm{Gaussian}(x, \mu, \sigma)	& \exp[-(x - \mu)^2 / (2 \sigma^2)] / (\sigma \sqrt{2\pi})	\nonumber\\
&\textrm{(also known as the Normal density)}\nonumber\\
&\textrm{LogNormal}(x, \mu, \sigma)	& \exp[-(\ln x - \mu)^2 / (2 \sigma^2)] / (x \sigma \sqrt{2\pi})
\nonumber\\
& \textrm{Chisq}(x, n)			& x^{n/2 -1} \exp(-x /2) / [2^{n/2} \Gamma(n/2)]	\nonumber\\
& \textrm{Gamma}(x, a, b)			& x^{b -1} a^b \exp(- a x) / \Gamma(b)	\nonumber\\
&\textrm{Exp}(x, a)				& a \exp(- a x) \nonumber\\
&\textrm{Beta}(x, n, m)			& \frac{\Gamma(n+m)}{\Gamma(n) \, \Gamma(m)}
x^{n-1} \, (1 - x)^{m-1} 
\label{eq:dist}
\end{align}
Particle physicists tend to use the term probability distribution for both discrete and 
continuous functions, such as the Poisson and Gaussian distributions, respectively. But, strictly speaking, the continuous functions are probability \emph{densities}, not probability
distributions. In order to compute a probability from a density we need to integrate the density 
over a finite set in $x$.

\subsection{Likelihood}
\label{sec:likelihood}
Let us assume that $p(x|\theta)$ is a \textbf{probability density function} (pdf) such that
$P(A| \theta) = \int_A p(x|\theta) \, dx$ is the probability of the statement  $A = x \in R_x$,
where $x$ denotes possible data, $\theta$
the parameters that
characterize the \textbf{probability model} (that is the probability together with all the assumptions on which it is based), and $R_x$ is a finite set. We shall use probability model
as shorthand for probability density function (for continuous variables) or
probability mass function (pmf) (basically, probabilities for discrete variables). If $x$ is discrete, then both
$p(x|\theta)$ and $P(A|\theta)$ are probabilities. The \textbf{likelihood function} is
simply the probability model $p(x|\theta)$ evaluated at the data $x_O$ actually obtained, i.e., the function $p(x_O|\theta)$.
The following are examples of likelihoods.

\begin{quote}
\paragraph*{Example 1}
In 1995, CDF and D\O\ discovered the top quark~\cite{Abe:1995hr, Abachi:1995iq} at Fermilab. The D\O\
Collaboration found  $x = N = 17$ events. For a counting experiment, the datum can be modeled using
\begin{align*}
	p(x | n) 	& = \textrm{Poisson}(x,  n) \quad \textrm{probability to get $x$ events}
\\ 	p(N | n) 	& = \textrm{Poisson}(N, n) \quad \textrm{likelihood of $N$ events}
\\ 			& = n^{N} \exp(-n) /  N!
\end{align*}
We shall analyze this example in detail in Lectures 2 and 3.

%
%
%

\paragraph*{Example 2}
Figure~\ref{fig:type1a} shows a plot of the distance modulus versus redshift for
$N = 580$ Type 1a supernovae~\cite{Suzuki:2011hu}. These heteroscedastic data\footnote{Data in which each item, $x_i$, or group of items has a different uncertainty.} $D = \{z_i, x_i \pm \sigma_i \}$ are modeled
using the likelihood
\begin{align*}
	p(D | \Omega_M, \Omega_\Lambda, Q) 	& = \prod_{i=1}^N \textrm{Gaussian}(x_i, \mu_i, \sigma_i),
\end{align*}
which is an example of an \emph{un-binned} likelihood. The cosmological model is
encoded in the distance modulus function $\mu_i$, which depends on the redshift $z_i$
and the matter density and cosmological constant parameters $\Omega_M$ and $\Omega_\Lambda$, respectively. (See Ref.~\cite{Dungan:2009fp} for an accessible introduction to the analysis of these data.)

\begin{figure}
\centering\includegraphics[width=0.5\textwidth]{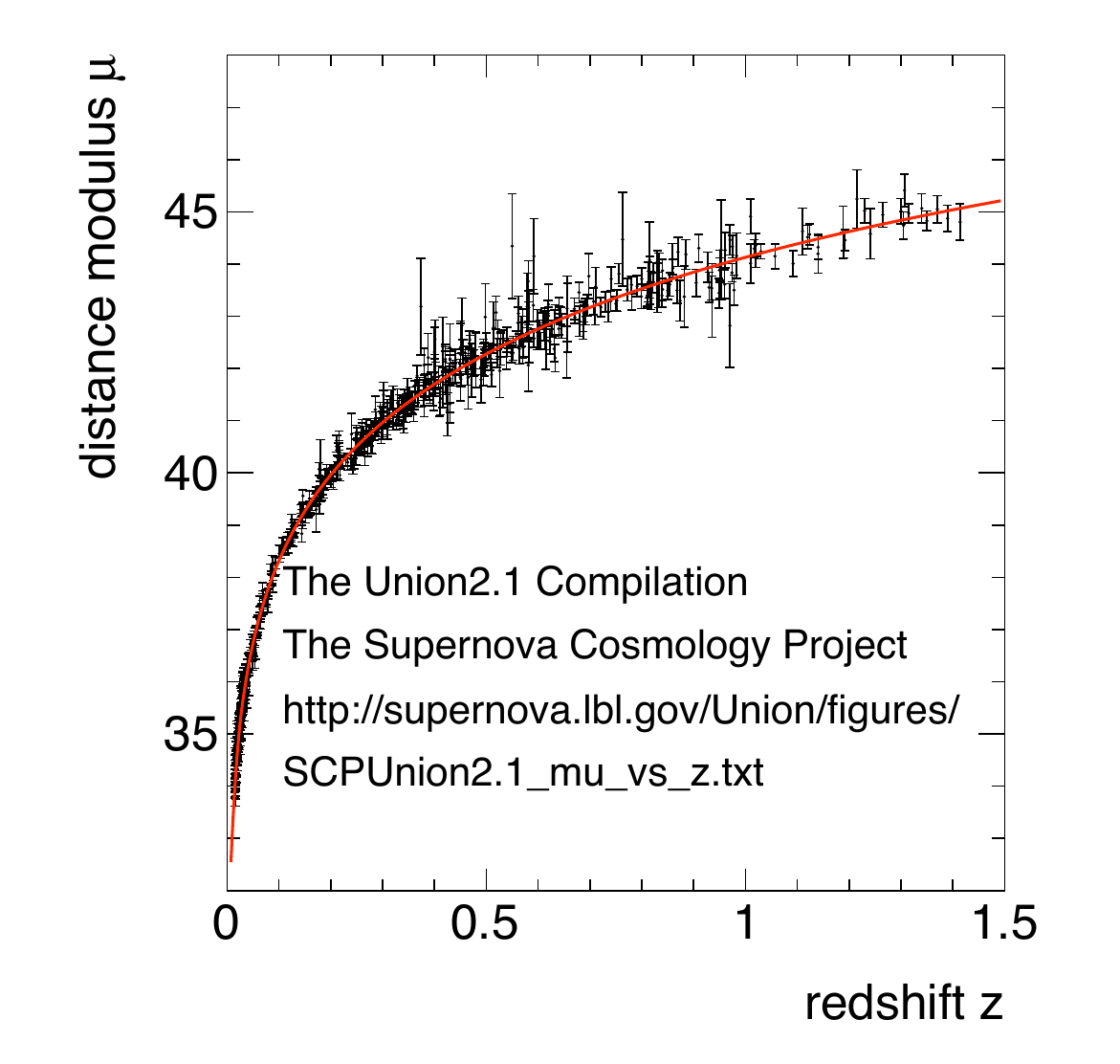}
\caption{Plot of the data points $(z_i, x_i \pm \sigma_i)$ for 580 Type 1a supernovae~\cite{Suzuki:2011hu} showing a fit of the standard cosmological model (with a cosmological constant) to
these data (curve).} 
\label{fig:type1a}
\end{figure}

\paragraph*{Example 3}
The discovery of a Higgs boson by ATLAS~\cite{Aad:2012tfa} and CMS~\cite{Chatrchyan:2012ufa} in the di-photon final state ($p p \rightarrow H \rightarrow \gamma\gamma$)
made use of an un-binned likelihood of the form,
\begin{align*}
	p(x| s, m, w, b) & = \exp[-(s + b)] \prod_{i=1}^N [ s f_s(x_i | m, w) + b f_b(x_i) ] 
	\\
	\textrm{where } 	x 	& = \textrm{di-photon masses} \\
				m	& = \textrm{mass of boson}	\\
				w	& = \textrm{width of resonance}	\\
				s	& = \textrm{expected (i.e., mean) signal count}	\\
				b	& = \textrm{expected background count}	\\
				f_s	& = \textrm{signal probability density}	\\
				f_b	& = \textrm{background probability density} \\ \\
				&  \framebox{\parbox{0.5\textwidth}{\textbf{Exercise 6:} Show that a binned multi-Poisson\\ likelihood yields an un-binned likelihood of this\\ form as the bin widths go to zero}}
\end{align*}
\end{quote}

\bigskip

The likelihood function is arguably the most important quantity in a statistical analysis
because it can be used to answer questions such as the following.
\begin{enumerate}
	\item How do I estimate a parameter?
	\item How do I quantify its accuracy?
	\item How do I test an hypothesis?
	\item How do I quantify the significance of a result?
\end{enumerate}
Writing down the likelihood function requires:
\begin{enumerate}
	\item identifying all that is \emph{known}, e.g., the observations,
	\item identifying all that is \emph{unknown}, e.g., the parameters,
	\item constructing a probability model for \emph{both}.
\end{enumerate}
Many analyses in particle physics do not use likelihood functions explicitly. However,
since the data we use are stochastic, the failure to reflect deeply on their
probabilistic nature and to model them explicitly leads to analyses that may not be as good as they could be. Deconstructing carefully what is being done in an
analysis is a habit that should be encouraged so that an accurate probabilistic model of the analysis can be constructed.

%
 
\section{Lecture 2:  the frequentist approach}
In this lecture, we consider statistical inference from the
frequentist viewpoint. In lecture 3, we consider the Bayesian approach.
In our opinion, both are needed to make sense of statistical inference,  though this
is not the dominant opinion in particle physics. 

The most important principle in the frequentist approach is that enunciated by 
the Polish statistician Jerzy Neyman in the 1930s, namely, 
\begin{quote}
\textbf{The frequentist principle}

The goal of a frequentist analysis is to construct statements so that a fraction $f \geq p$ of
them are guaranteed to be true over an infinite ensemble of statements.
\end{quote}
The fraction $f$ is called the \textbf{coverage probability},  or coverage for short, and $p$ is called the \textbf{confidence level} (C.L.). A procedure which satisfies the frequentist principle is said to \emph{cover}.
The confidence level as well as the coverage is a property of the
ensemble of statements. Consequently, the confidence level may change if the ensemble changes. In a seminal paper published in 1937, Neyman~\cite{Neyman37} 
invented the concept of the confidence interval, a way to quantify
uncertainty, that respects the frequentist principle. The confidence interval is such an important idea that it is worth working through the concept in detail.

\subsection{Confidence intervals}
\begin{figure}
\centering\includegraphics[width=0.8\textwidth]{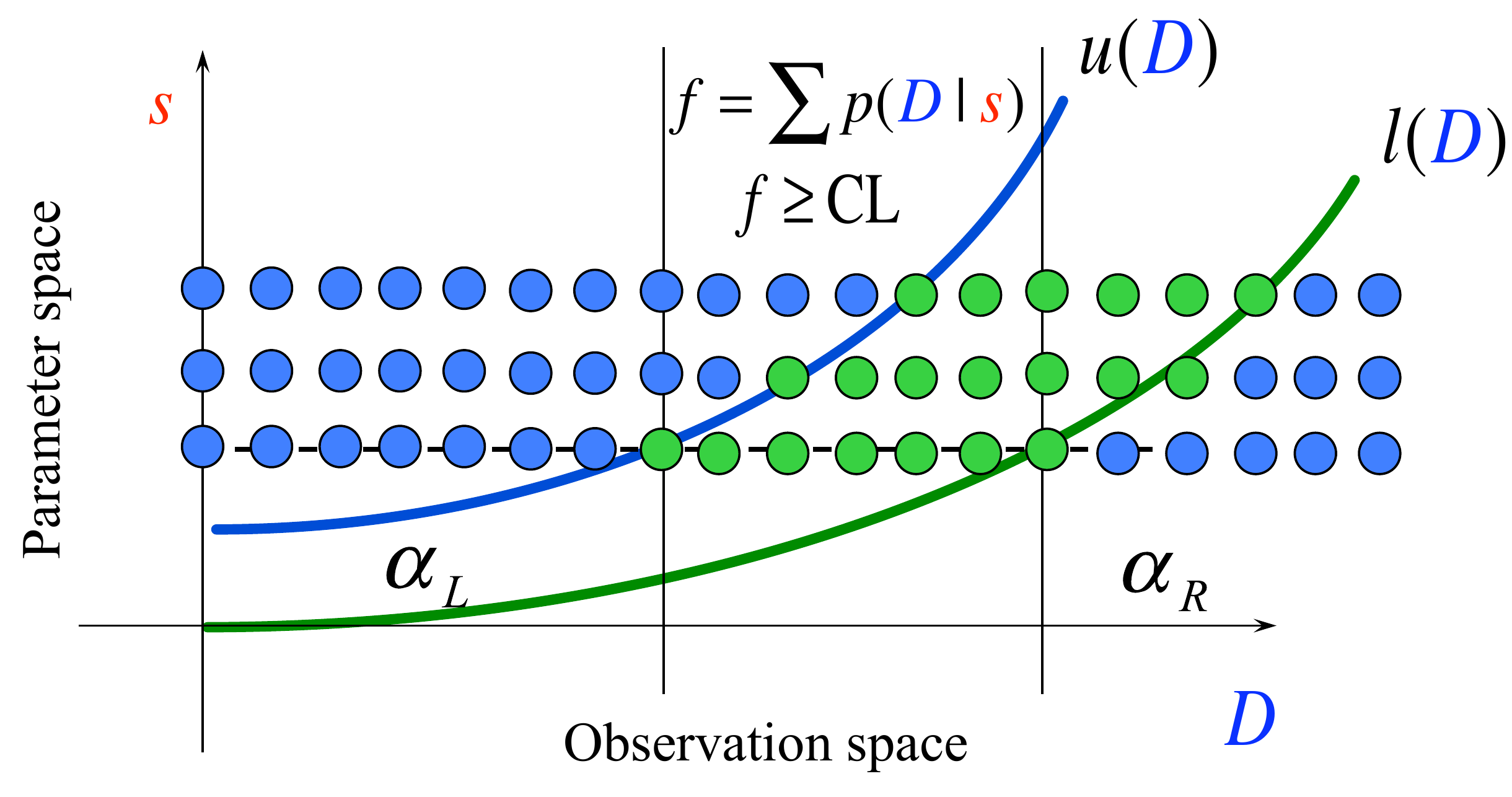}
\caption{The Neyman construction. Plotted is the Cartesian product of the parameter
space, with parameter $s$, and the space of observations with potential observations $D$. 
For a given value of $s$, the observation space is partitioned into three disjoint intervals,
such that the probability to observe a count $D$ within the interval demarcated by
the two vertical lines is $f \geq p$, where p = C.L. is the desired confidence level. The
inequality is needed because, for discrete data, it may not be possible to find an interval
with $f = p$ exactly.} 
\label{fig:neyman}
\end{figure}

Consider an experiment
that observes $D$ events with expected (that is, mean) signal $s$ and no background. Neyman devised a way to make statements of the form
\begin{align}
	s \in [ l(D), \, u(D) ],
\end{align}
with the \emph{a priori} guarantee that at least a fraction $p$ of them will be true
over an ensemble of statements of this kind. A procedure for constructing such
intervals is called a \textbf{Neyman construction}. The frequentist principle 
must hold for \emph{any} ensemble of experiments, not necessarily all making the same kind of
observations and statements. For simplicity, however, we shall consider the
experiments  to be of the same kind and to be completely specified by a single unknown
parameter $s$. The
Neyman construction is illustrated in Fig.~\ref{fig:neyman}. 

The construction proceeds as follows. Choose a value of $s$ and use some rule to find
an interval in the space of observations (or, more generally, a region), for example, the
interval defined by the two vertical lines in the center of the figure, such that the probability to obtain a count in this interval is  $f \geq p$, where $p$ is the desired confidence level. Then move to another
value of $s$ and repeat the procedure. The procedure is repeated for a sufficiently dense set of points in the parameter space over a sufficiently large range in $s$. When this is done, as illustrated in Fig.~\ref{fig:neyman}, the intervals of probability content
$f$ will form a band in the Cartesian product of the parameter space and the observation space.
The upper edge of this band defines the curve $u(D)$, while the lower edge defines the curve
$l(D)$. These curves are the outcome of the Neyman construction.

For a given value of  $s$, the interval with probability content $f$ in the space
of observations
is not unique since different rules for choosing the interval will, in general, yield different intervals. Neyman suggested choosing the interval so that the probability to obtain an observation 
to the right or left of
the interval are the same (for a given value of $s$), which  yields the so-called \textbf{central intervals}. 
One virtue of these intervals is that their boundaries can be more efficiently calculated by
solving the equations,
\begin{align}
	P(x \leq   D | u) & = \alpha_L, \nonumber\\
	P(x \geq D |  l)  & = \alpha_R,
\end{align}
a mathematical fact that becomes clear if we stare at Fig.~\ref{fig:neyman} long enough.

Another rule was suggested by Feldman and Cousins~\cite{FC}. For our example, the Feldman-Cousins
rule requires that the potential observations $\{D\}$ be ordered in descending order, $D_{(1)}, D_{(2)}, \cdots$, of the likelihood ratio $p(D | s) / p(D | \hat{s})$, where
$\hat{s}$ is the maximum likelihood estimator (see Sec.~\ref{sec:profile}) of the parameter $s$.
Once ordered, we compute the running sum $f = \sum_j p(D_{(j)} | s)$ until $f$ equals or just exceeds the desired
confidence level $p$. This rules does not guarantee that the potential observations $D$ are
contiguous, but this does not matter because we simply take the minimum element of the set $\{ D_{(j)} \}$ to be
the lower bound of the interval and its maximum element to be the upper bound. 

\begin{wrapfigure}{R}{0.5\textwidth}
\centering\includegraphics[width=0.5\textwidth]{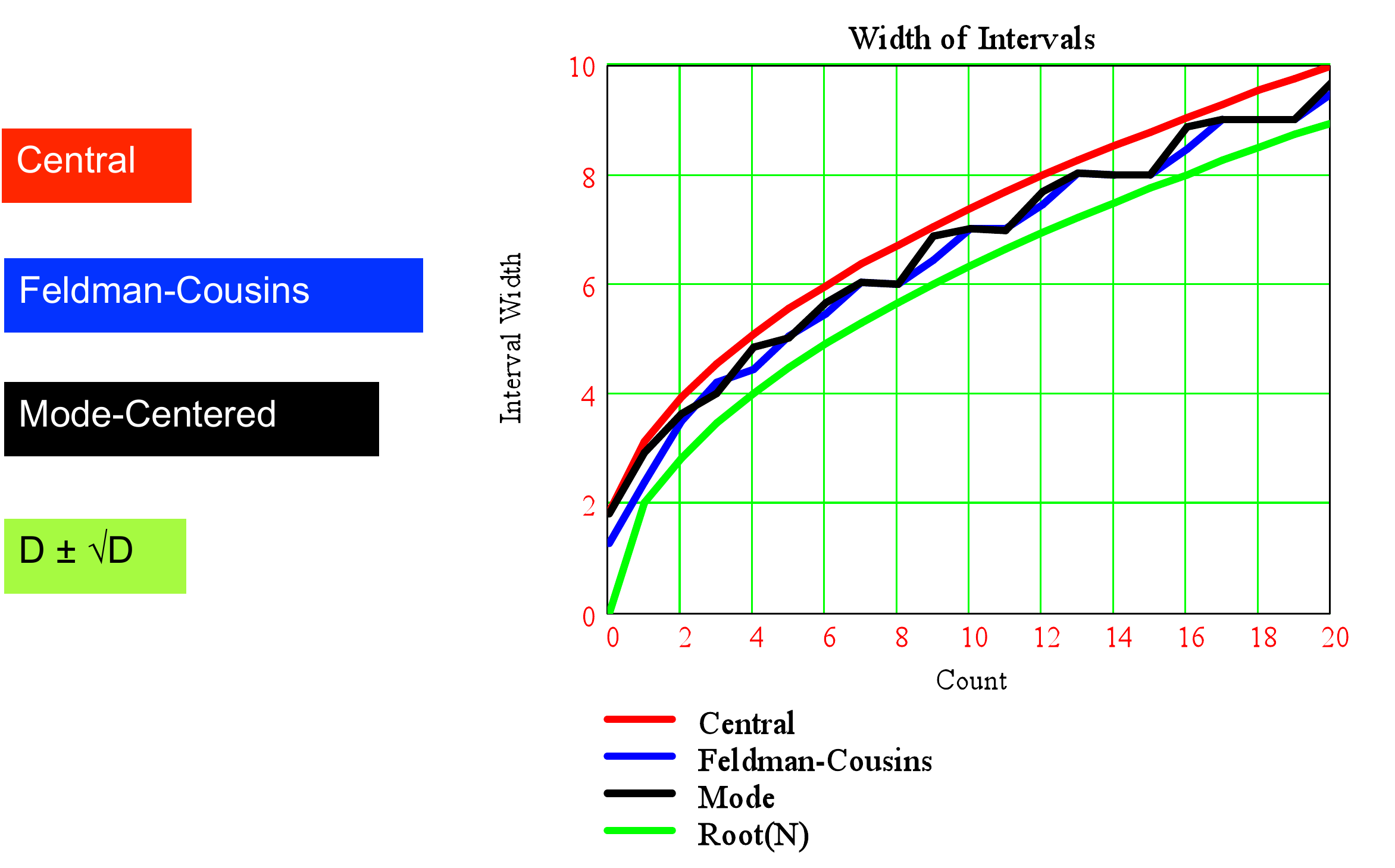}
\caption{Interval widths as a function of count $D$ for four sets of intervals.} 
\label{fig:ciwidths}
\end{wrapfigure}

Another simple rule is the mode-centered rule: order $D$ in descending order of $p(D| s)$ and 
proceed as with the Feldman-Cousins rule. 
In principle, absent criteria for choosing a rule, there is nothing
to prevent the use of \emph{ordering rules} randomly chosen for different values of $s$! 
Figure~\ref{fig:ciwidths} compares the widths of the
intervals $[l(D), u(D)]$ for three different ordering rules, central, Feldman-Cousins, and mode-centered as a function of the count $D$. It is instructive to compare these widths with those provided by
the well-known root(N) interval, $l(D) = D - \sqrt{D}$ and $u(D) = D + \sqrt{D}$. Of the three sets of intervals, the ones suggested by Neyman are the widest, the Feldman-Cousins and mode-centered ones are of similar width, while the root(N) intervals are the shortest. So why are we going through
all the trouble of the Neyman construction? We shall return to this question shortly.

\begin{wrapfigure}{L}{0.5\textwidth}
\centering\includegraphics[width=0.5\textwidth]{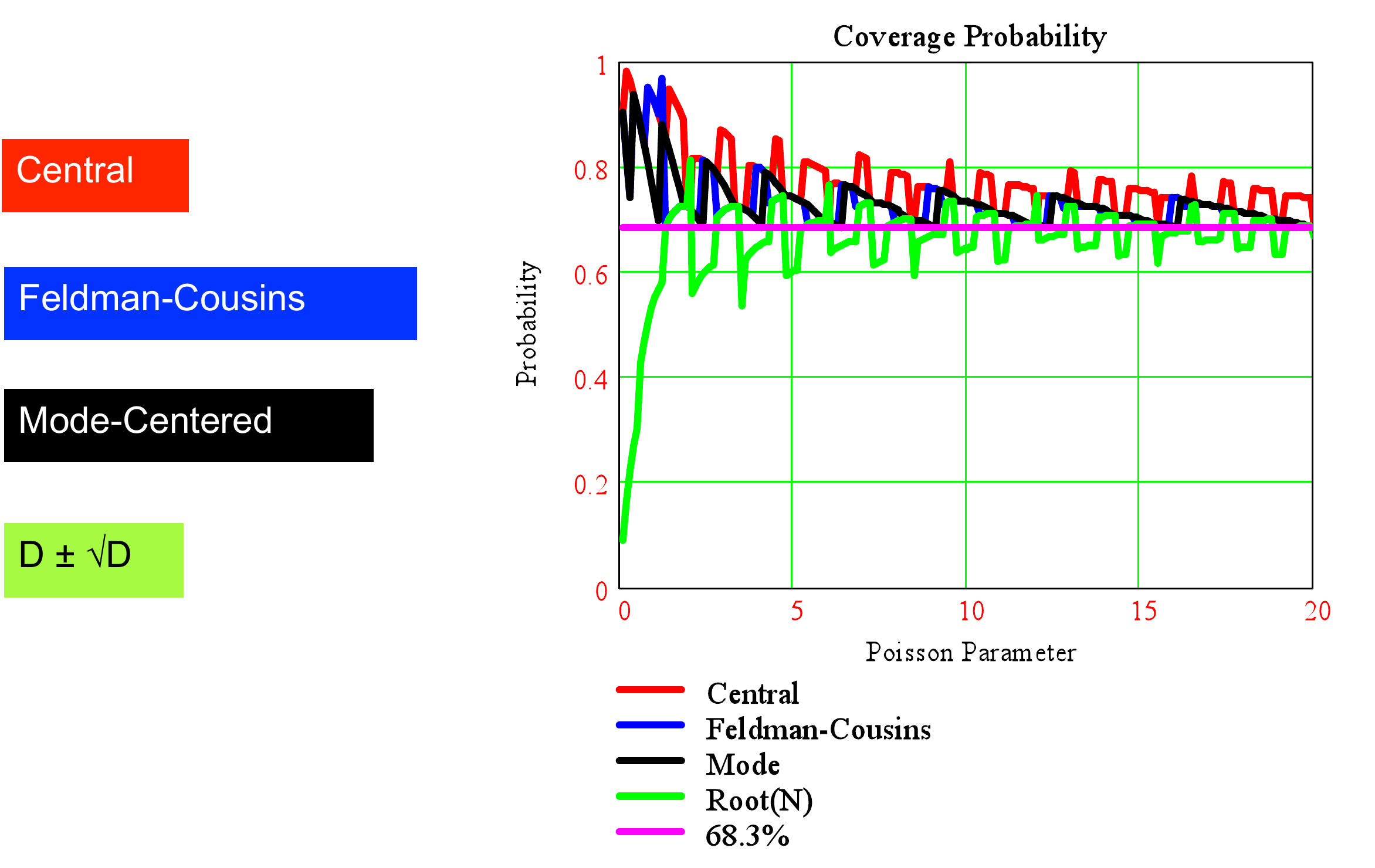}
\caption{Interval widths as a function of count $D$ for four sets of intervals.} 
\label{fig:coverage}
\end{wrapfigure}

Having completed the Neyman construction and found the curves $u(D)$ and $l(D)$ 
we can use the latter to make statements of
the form $s \in [l(D), \, u(D)]$: for a given observation $D$, we simply read off
the interval $[l(D), u(D)]$ from the curves. For example, suppose in Fig.~\ref{fig:neyman} that the true value of $s$ is
represented 
by the horizontal line that intersects  the curves $u(D)$ and $l(D)$ and which therefore defines
the interval demarcated by the two vertical lines. If the observation $D$ happens to fall in the interval to the left of the left vertical line, or to the right of the right vertical line, then the interval
$[l(D), \, u(D)]$ will not bracket $s$. However, if $D$ falls between the two vertical
lines, the interval  $[l(D), \, u(D)]$ will bracket $s$. Moreover, by virtue of the Neyman construction, a fraction $f$ of the intervals $[l(D), \, u(D)]$ will bracket the value of $s$ whatever its value happens to be, which brings us back to the question about the root(N) intervals. Figure~\ref{fig:coverage} shows the coverage probability over the parameter space of $s$. As expected,
the three rules, Neyman's, that of Feldman-Cousins, and the mode-centered, satisfy the condition coverage probability $\geq$ confidence level over all values of $s$ that are
possible \emph{a priori}; that is, the intervals cover. However, the root(N) intervals do not and indeed fail badly for $ s < 2$.

However, the coverage probability of the  root(N) intervals bounces around the (68\%) confidence level for vaues of $s > 2$. Therefore, if we knew for sure that $s > 2$, it would
seem that using the root(N) intervals may not be that bad after all. 


So what, after all this, does the statement
$s \in  [l(D), \, u(D)]$ at $100 p$\% C.L. mean in this approach, given that  $p$ is a property of the ensemble to which this 
statement belongs? In means this: $s \in  [l(D), \, u(D)]$ is a member of an ensemble of statements a fraction $f \geq p$ of which are true. 
In principle, in order to verify this we need just count how many
statements of the form $s \in  [l(D), \, u(D)]$ are true and divide by the total number of
statements. Unfortunately, this requires that we know which statements are true.
But if we knew that we would not need a theory of statistical inference!  

Neyman required a procedure to cover whatever the value of \emph{all} the parameters, 
be they known or unknown, of the probability models that describe the data generation
mechanisms.
This is a very tall order, which cannot be met in general. In practice, we resort to
approximations,  the most widely used of which is the profile likelihood to which we now turn.

\subsection{The profile likelihood}
\label{sec:profile}
As noted in Section~\ref{sec:likelihood}, likelihood functions can be used to estimate the parameters
on which they depend. The method of choice to do so, in a frequentist analysis, is called \textbf{maximum
likelihood}, a method first used by Karl Frederick Gauss and developed into a formidable statistical tool in the 1930s by 
Sir Ronald A. Fisher~\cite{Fisher}, perhaps the most influential statistician of the twentieth century. The D\O\ top quark
discovery example illustrates the
method. 

\begin{quote}
\paragraph*{Example: Top Quark Discovery Revisited}
We start by listing
\begin{align*}
& \textbf{the knowns} \\
	& D = N, B \text{ where} \\
	& N = 17 \textrm{ observed events} \\
	& B = 3.8 \textrm{ estimated background events with uncertainty } \delta B = 0.6 \\
&\textbf{and the unknowns} \\
	& b \quad\textrm{mean background count}\\
	& s \quad\textrm{mean signal count}.
\end{align*} 
Next, we construct a probability model for the data $D = N, B$. Since this is a counting
experiment, we shall assume that $p(x| s, b)$ includes a Poisson distribution with mean
count $s + b$. In the absence of details about how the background $B$ was arrived
at, the standard assumption is that data of the form $y \pm \delta y$ can be modeled
with a Gaussian (or normal) density. However,  we shall do something slightly better.

Suppose that the observed count in the control region is $Q$ and the mean count is $b k$, where $k$ (ideally) is the known scale factor between the control and signal regions.  But, since
we are given $B$ and $\delta B$ rather than $Q$ and $k$, we need to relate the two pairs of numbers.  
The simplest model  is $B = Q / k$ and $\delta B = \sqrt{Q} / k$ from which we can infer an effective count $Q$ using $Q = (B / \delta B)^2$. Since the scale factor $k$
is not given, we shall use the obvious estimate $k \sim Q / B = B / \delta B^2$.
With these assumptions, our likelihood function is
\begin{eqnarray}
	\label{eq:toplh}
	p(D | s, b) 	& = & \textrm{Poisson}(N, s + b) \, \textrm{Poisson}(Q, bk), \\
	\textrm{where} \nonumber\\
		Q 	& =  & (B / \delta B)^2  = 41.11,\nonumber\\
		k 	& = & B / \delta B^2  = 10.56. \nonumber
\end{eqnarray}
The first term in Eq.~(\ref{eq:toplh}) is the likelihood for the count $N = 17$, while
the second term is the likelihood for $B = 3.8$, or equivalently the count $Q$.   The
fact that $Q$ is not an integer causes no difficulty; we merely continue
 the Poisson distribution to non-integer $Q$ using
$(bk)^Q \exp(-bk) / \Gamma(Q+1)$.

The maximum likelihood estimators 
for $s$ and $b$ are found by maximizing Eq.~(\ref{eq:toplh}), that is, by solving the equations 
\begin{align}
	\frac{\partial \ln p(D|s, b)}{\partial s} & = 0\quad\textrm{leading to } \hat{s} = N - B, \nonumber\\ 
\textrm{and }	\frac{\partial \ln p(D|s, b)}{\partial b} & = 0\quad\textrm{leading to } \hat{b} = B, \nonumber
\end{align}
as expected.

A more complete model would account for the uncertainty in
$k$. 
\end{quote}
\bigskip

The maximum likelihood method is the most widely used method for
estimating parameters because it generally leads to reasonable estimates. But the
method has features, or encourages practices, which, somewhat uncharitably, we label the
good, the bad, and the ugly!
\begin{itemize}
	\item \emph{The Good}
	\begin{itemize}
		\item Maximum likelihood estimators are consistent: the RMS goes to zero as more
		and more data are included in the likelihood. This is an extremely important property,
		which basically says it makes sense to take more data because we shall get more 			accurate results. One would not knowingly use an inconsistent estimator!
		
		\item If an unbiased estimator for a parameter exists the maximum
		likelihood method will find it.
		
		\item Given the MLE for $s$, the MLE for any function $y = g(s)$ of $s$ is,
		very conveniently, just $\hat{y} = g(\hat{s})$. This is a very nice practical feature which
		makes it possible to maximize the likelihood using the most convenient parameterization of it and then transform back to the parameter of interest at the end. 
	\end{itemize}
	
	\item \emph{The Bad (according to some!)}
	\begin{itemize}
		\item In general, MLEs are biased.\\ \\
		\framebox{\parbox{0.5\textwidth}{\textbf{Exercise 7:} Show this\\
		Hint: Taylor expand $y = g(\hat{s} + h)$ about the MLE $\hat{s}$,\\
		then consider its ensemble average. }}		
	\end{itemize}

	\item \emph{The Ugly (according to some!)}
	\begin{itemize}
		\item The fact that most MLEs are biased encourages the routine application of bias correction, which can waste data and, sometimes, yield absurdities.
	\end{itemize}
\end{itemize}

\noindent
Here is an example of the seriously ugly.
\begin{quote}
\paragraph*{Example}
For a discrete probability distribution $p(k)$, the \textbf{moment generating function} is the ensemble average
\begin{align*}
	G(x) & = \langle e^{xk} \rangle \\
		& = \sum_{k} e^{xk} \, p(k).
\end{align*}
For the binomial, with parameters $p$ and $n$, this is
\begin{align*}
	G(x) & = (e^x p + 1 - p)^n, \quad \framebox{\textbf{Exercise 8a:} Show this}
\end{align*}
which is useful for calculating \textbf{moments}
\begin{align*}
	\mu_r & = \left. \frac{d^rG}{dx^r}\right |_{x=0} = \sum_k k^r \, p(k),
\end{align*}
e.g., $\mu_2 = (np)^2 + np - np^2$ for the binomial distribution.
Given that $k$ events out $n$ pass a set of cuts, the MLE of the event selection efficiency is
the obvious estimate $\hat{p} = k / n$. The equally obvious estimate of $p^2$ is $( k / n)^2$.
But,
\begin{align*}
	\langle ( k / n)^2 \rangle & = p^2 + V / n , \quad \framebox{\textbf{Exercise 8b:} Show this}
\end{align*}
so $(k / n)^2$ is a biased estimate of $p^2$ with positive bias $V / n$. The unbiased estimate of $p^2$ is
\begin{align*}
	 k(k-1) / [ n (n - 1)] , \quad \framebox{\textbf{Exercise 8c:} Show this}
\end{align*}
which, for a single success, i.e., $k = 1$, yields the sensible estimate $\hat{p} = 1 / n$, but
the less than useful $\hat{p^2} = 0!$
\end{quote}


\bigskip

In order to infer a value for the parameter of interest, for example, 
the signal $s$ in
our  2-parameter likelihood function in Eq.~(\ref{eq:toplh}), the likelihood
must be reduced to one involving the parameter of interest only, here $s$, 
by getting rid of all the \textbf{nuisance} parameters, here the background
parameter $b$. A nuisance parameter is  any parameter that is not of current interest.
In a strict frequentist calculation, this reduction to the parameter of interest must be done
in such a way as to respect the frequentist principle: \emph{coverage probability $\geq$ confidence level}. In general, this is very difficult to do exactly.

In practice, we replace all nuisance parameters by their \textbf{conditional maximum likelihood
estimates} (CMLE). The CMLE is the maximum likelihood estimate conditional on
a \emph{given} value of the current parameter (or parameters) of interest. In the top discovery example, we construct an estimator of $b$ as a function of $s$, $\hat{b}(s)$, and
replace $b$  in the likelihood $p(D | s, b)$ by $\hat{b}(s)$ to yield a function
$p_{PL}(D | s)$ called the \textbf{profile likelihood}.
\begin{quote}
\emph{Since the profile likelihood entails an approximation, namely, replacing unknown parameters by their conditional estimates, it is no longer the likelihood but rather an approximation to it. Consequently, 
the frequentist principle is not guaranteed to be satisfied exactly.}
\end{quote}
But, if certain conditions are met (Wilks' theorem, 1938), roughly that the
MLEs do not occur on the boundary of the parameter space and the likelihood becomes
ever more Gaussian as the data become more numerous --- that is, in the so-called
\textbf{asymptotic limit}, then if the true density of $x$ is $p(x| s, b)$ the random number
\begin{align}
	t(x, s) & = -2 \ln \lambda(x, s), \\
	\textrm{where } \lambda(x, s)  & = \frac{p_{PL}(x | s)}{ p_{PL}(x | \hat{s})}, 
	\label{eq:wilks}
\end{align}
has a probability density that converges to a $\chi^2$ density with one degree of
freedom. More generally, if the numerator of $\lambda$ contains $m$ free parameters the
asymptotic density of $t$ is a $\chi^2$ density with $m$ degrees of freedom. Therefore, we may take $t(D, s)$ to be a $\chi^2$ variate, at least
approximately, and solve $t(D, s) = n^2$ for $s$ to get 
approximate $n$-standard deviation confidence intervals. In particular, if we solve $t(D, s) = 1$, we
obtain approximate 68\% intervals. This calculation is what {\tt Minuit}, and now {\tt TMinuit}, has done countless times since the 1970s! 
Wilks' theorem provides the main justification for using the profile likelihood. 
We again use the top discovery example to illustrate the procedure.

\begin{quote} 
\paragraph*{Example: Top Quark Discovery Revisited Again}
The conditional MLE of $b$ is found to be
\begin{align}
	\hat{b}(s) & = \frac{g + \sqrt{g^2 + 4 (1 + k) Q s}}{2(1+k)}, \\
	\textrm{where} \nonumber\\
	g & = N + Q - (1+k) s.\nonumber
	\label{eq:bhat}
\end{align}

%

\begin{figure}
\centering\includegraphics[width=0.48\textwidth]{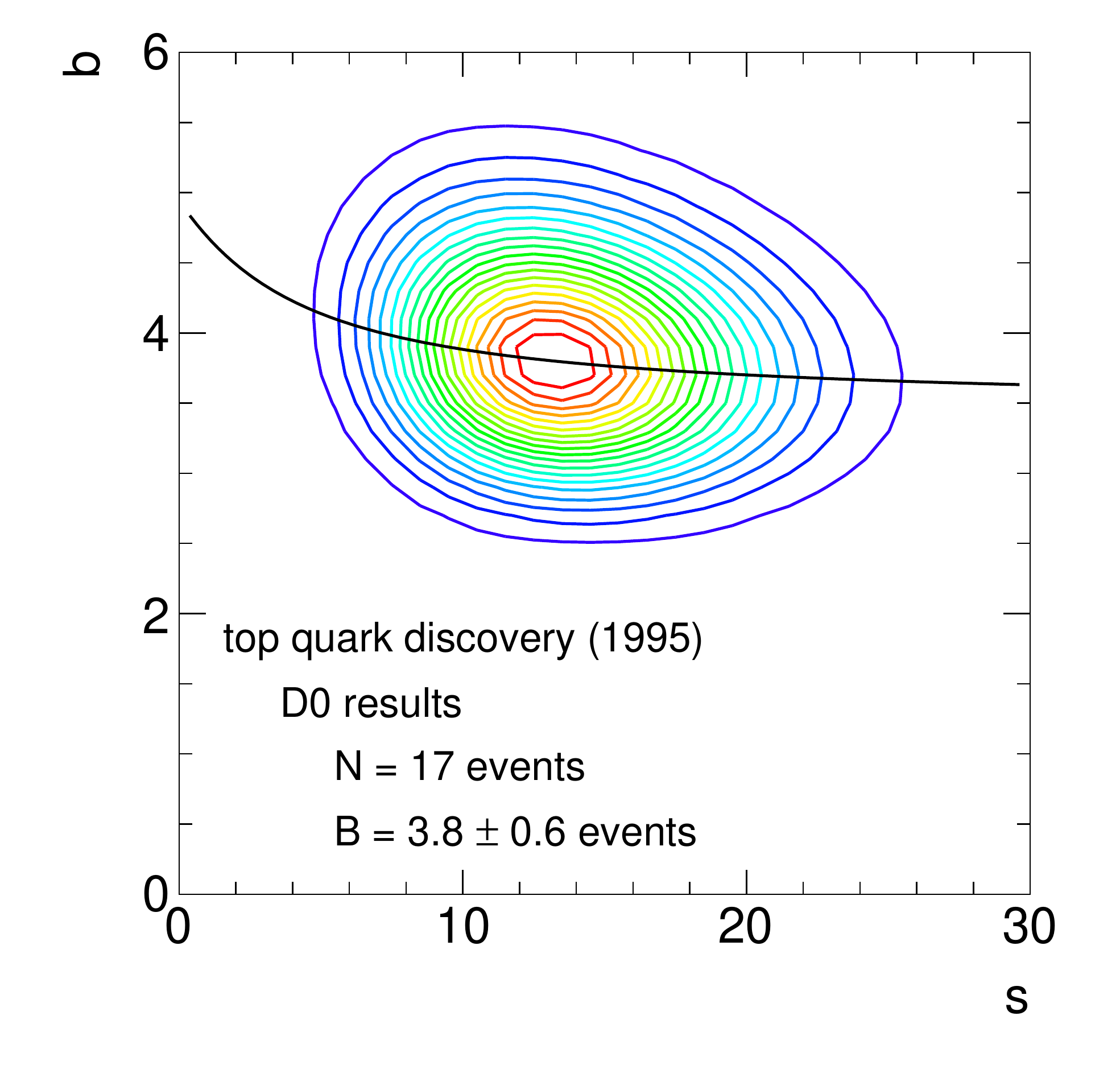}
\centering\includegraphics[width=0.48\textwidth]{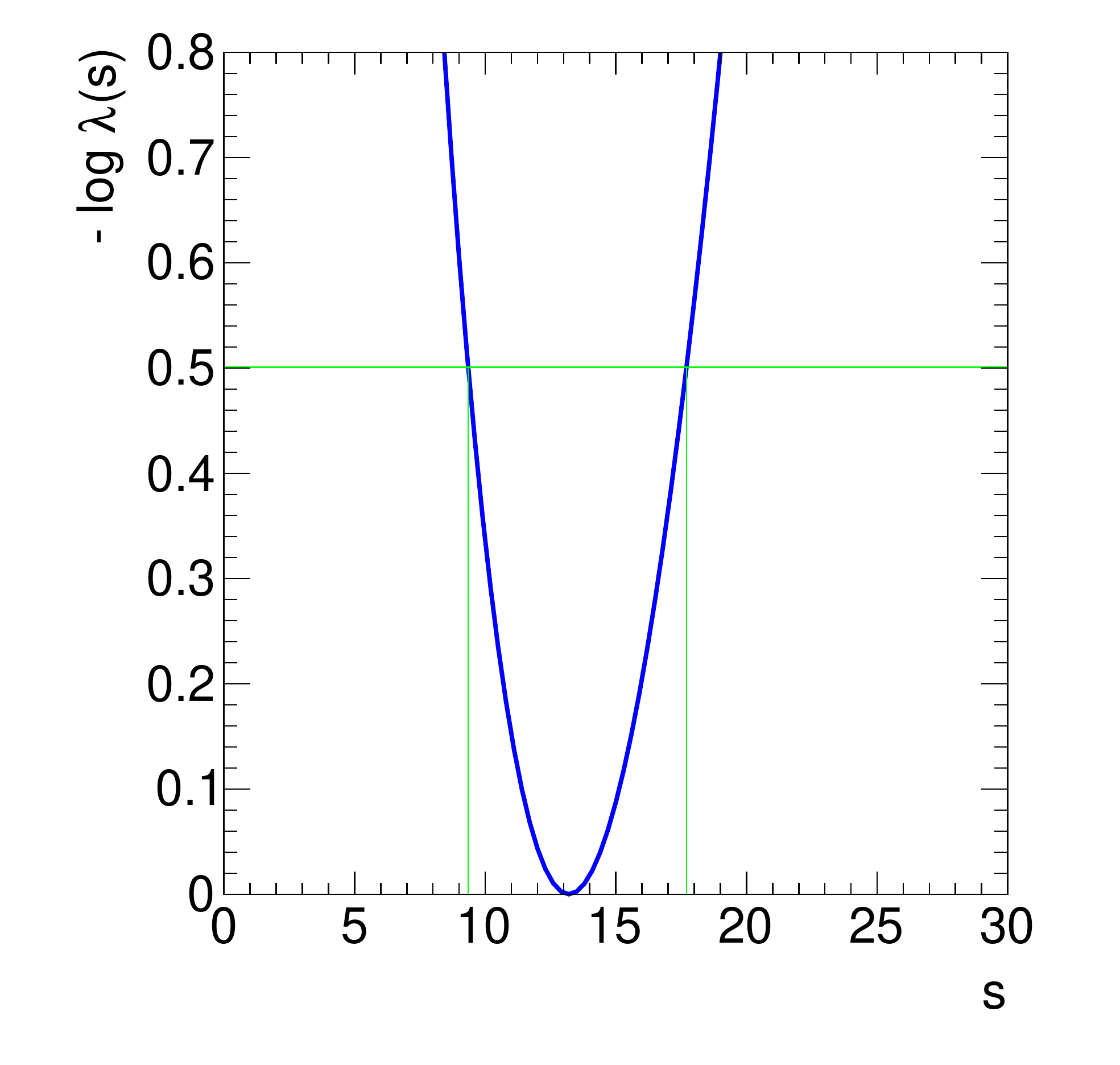}
\caption{(a) Contours of the D\O\ top discovery likelihood
and the graph of $\hat{b}(s)$.
(b) Plot of $-\ln \lambda(17, s)$ versus the expected signal $s$. The vertical lines show the boundaries of the approximate 68\% interval.}
\label{fig:toppl}
\end{figure}

The likelihood $p(D | s, b)$ is shown in Fig.~\ref{fig:toppl}(a) together with
the graph of $\hat{b}(s)$. The mode (i.e. the peak) occurs at $s = \hat{s} = N - B$.
By solving $$-2 \ln \frac{p_{PL}(17 | s)}{ p_{PL}(17 | 17 - 3.8)}  = 1$$ for $s$ we get two solutions
$s = 9.4$ and $s = 17.7$. Therefore, we can make the statement
$s \in [9.4, 17.7]$ at approximately 68\% C.L.  Figure~\ref{fig:toppl}(b) shows a plot of 
$-\ln \lambda(17, s)$ created using 
the {\tt RooFit}~\cite{RooFit} and {\tt RooStats}~\cite{RooStats} packages.

\smallskip

\framebox{\textbf{Exercise 9:} Verify this interval using the {\tt RooFit/RooStats} package}

\medskip

Intervals constructed this way are not guaranteed to
satisfy the frequentist principle. In practice, however, 
their coverage is very good for the typical probability models
used in particle physics, even for modest amounts of
data. 
\end{quote}
%
%
\bigskip

\subsection{Hypothesis tests}
It is hardly possible in experimental particle physics to avoid testing hypotheses, testing
 that invariably leads to decisions. For example, electron identification entails hypothesis testing; given data
$D$ we ask: is this particle an isolated electron or is it not an isolated electron? 
In the discovery of the Higgs boson, we had to test whether, given the data available in early summer 2012,  the Standard Model without a Higgs boson, a somewhat ill-founded background-only model, or
the Standard Model with the boson of July 2012, the background $+$ signal model, was
the preferred hypothesis. We decided that the latter model was preferred and announced the
discovery of a new boson. Given the ubiquity of hypothesis testing, it is important to have
a grasp of the methods that have been invented to implement it.

One method was due to Fisher~\cite{Fisher}, another was
invented by Neyman, and a third (Bayesian) method was proposed by 
Sir Harold Jeffreys~\cite{Jeffreys}, all around the same time.
 We first describe the method of Fisher, then follow with a description of the method of
Neyman. For concreteness, we consider the problem of deciding between a background-only
model and a background $+$ signal model.

\begin{wrapfigure}{R}{0.5\textwidth}
	\centering\includegraphics[width=0.45\textwidth]{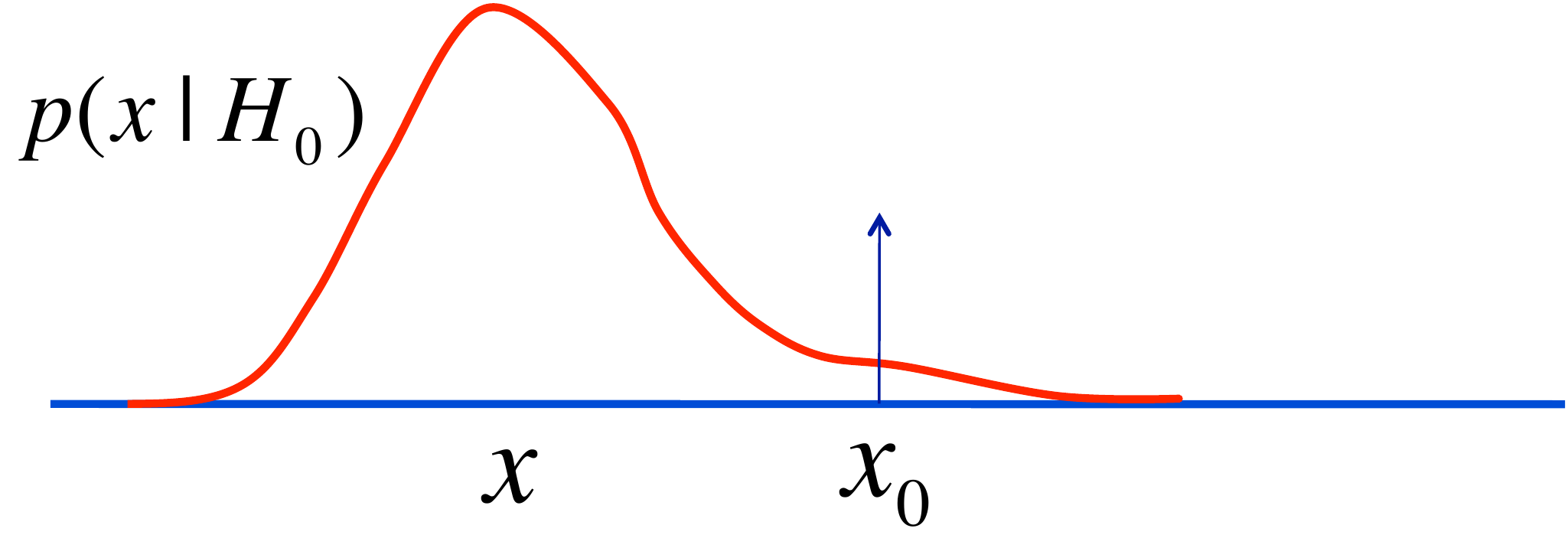}
	\caption{The p-value is the tail-probability,  $P(x > x_0| H_0)$, calculated from the probability density under
	the null hypothesis, $H_0$. Consequently, the probability density of
	the p-value under the null hypothesis is $\textrm{Uniform}(x, 1)$.}
	\label{fig:pvalue1}
\end{wrapfigure}

\bigskip

\subsubsection{Fisher's approach} In Fisher's approach, we construct a  \textbf{null hypothesis}, often denoted by $H_0$,
and \emph{reject} it should some measure  be judged
small enough to cast doubt on the validity of this  hypothesis. In our
example, the null hypothesis is the background-only model, for example, the SM without a Higgs boson. The measure is called a \textbf{p-value} and is defined by
\begin{align}
	\textrm{p-value}(x_0) = P( x > x_0| H_0), 
\end{align}
where $x$ is a statistic designed so that large values indicate
departure from the null hypothesis. This is illustrated in Fig.~\ref{fig:pvalue1}, which shows
the location of the observed value $x_0$ of $x$. The p-value is the probability that $x$ could
have been higher than the $x$ actually observed.
It is argued that a small p-value implies that either the null hypothesis is false or something rare has occurred. If 
the p-value is extremely
small, say $\sim 3 \times 10^{-7}$, then of the two possibilities the most common response
is to presume the null to be false. If we apply this method to the D\O\ top quark discovery data, and 
neglect the uncertainty in the null hypothesis, we find
\begin{align*}
	\textrm{p-value} & = \sum_{N=17}^\infty \textrm{Poisson}(N, 3.8) = 5.7 \times 10^{-7}.
\end{align*}
We usually report a more intuitive number by converting the p-value to the scale defined by 
\begin{align}
	Z & = \sqrt{2} \, \textrm{erf}^{-1}(1 - 2\textrm{p-value}).
\end{align}
This is the number of Gaussian standard deviations
away from the mean\footnote{$\textrm{erf}(x) = \frac{1}{\sqrt{\pi}} \int_{-x}^x \exp(-t^2) \, dt$ is the error funtion and $\textrm{erf}^{-1}(x)$ is its inverse.}. 
A p-value of $5.7 \times 10^{-7}$ corresponds to a $Z$ of $4.9\sigma$. The $Z$-value can be
calculated using the {\tt Root} function $$Z = \textrm{\tt -TMath::NormQuantile(p-value)}.$$

\begin{wrapfigure}{R}{0.5\textwidth}
	\centering\includegraphics[width=0.45\textwidth]{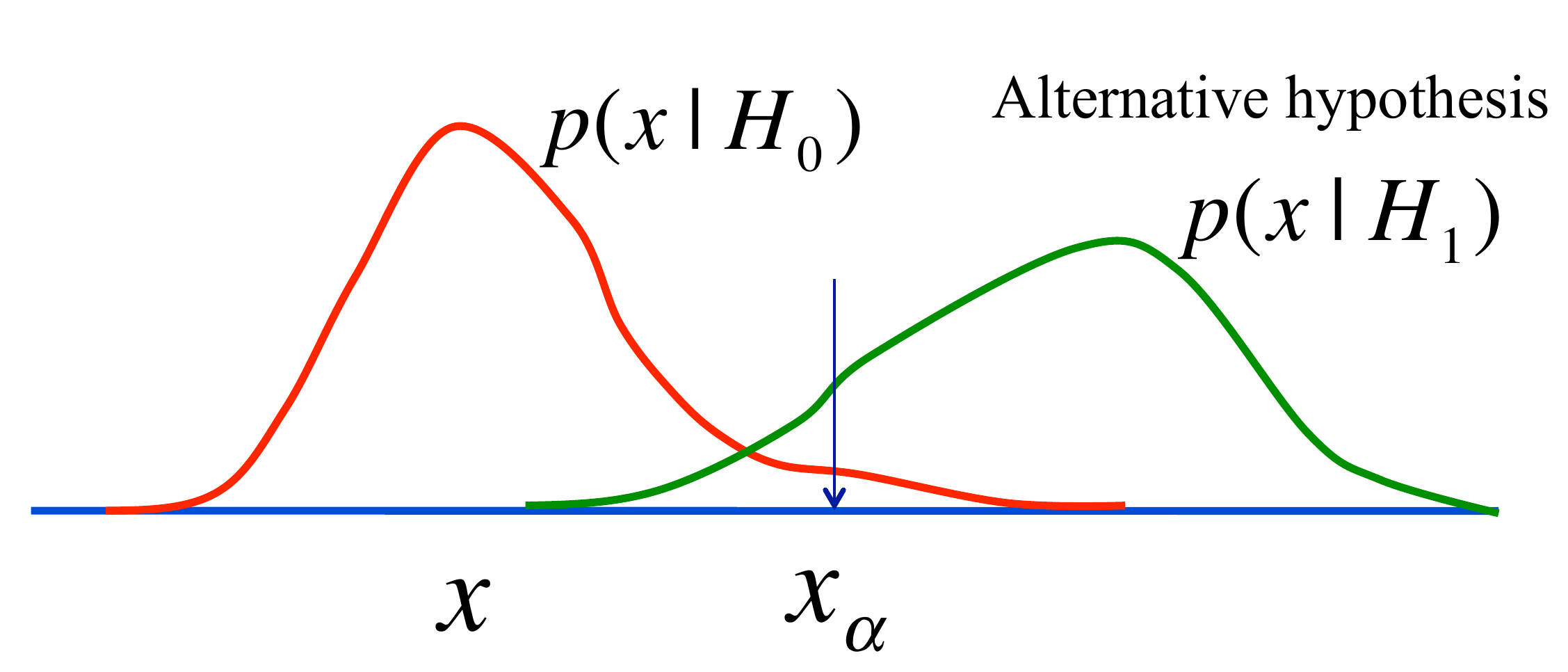}
	\caption{Distribution of a test statistic $x$ for two hypotheses, the null $H_0$ and the
	alternative $H_1$. In Neyman's approach to testing, $\alpha = P(x > x_\alpha|H_0)$ is a \emph{fixed}
	probability called the significance of the test, which for a given class of experiments corresponds the threshold $x_\alpha$. The hypothesis $H_0$ is rejected if $x > x_\alpha$.}
	\label{fig:neymantest1}
\end{wrapfigure}

\subsubsection{Neyman's approach}
In Neyman's approach \emph{two} hypotheses are considered, the null hypothesis $H_0$ and
an alternative hypothesis $H_1$. This is illustrated in Fig.~\ref{fig:neymantest1}. In our
example, the null is the same as before but the alternative hypothesis is the SM with a Higgs boson. 
Again, one generally chooses $x$ so that large values would cast doubt on 
the validity of $H_0$. However, the Neyman test is specifically designed to
respect the frequentist principle, which is done as follows. A \emph{fixed} probability $\alpha$ is
chosen called the significance (or size) of the test, which for a specific class of experiments corresponds to some threshold $x_\alpha$ defined by
\begin{align}
	\alpha & = P( x > x_\alpha | H_0).
\end{align}
 Should the observed value $x_0 > x_\alpha$, or
equivalently, p-value($x_0$) $< \alpha$, the hypothesis $H_0$ is rejected in favor of the
alternative. 
In 
particle physics, in addition to applying the Neyman hypothesis test, we also report the
p-value. This is sensible because there is a more information in the p-value than merely reporting the fact that a null hypothesis was rejected at a significance level of $\alpha$.  

The Neyman method satisfies the frequentist principle by construction. Since the significance of the test is fixed, $\alpha$ is the relative frequency with which \emph{true}
null hypotheses would be rejected and is called the \textbf{Type I} error rate. 

\begin{wrapfigure}{L}{0.5\textwidth}
	\centering\includegraphics[width=0.45\textwidth]{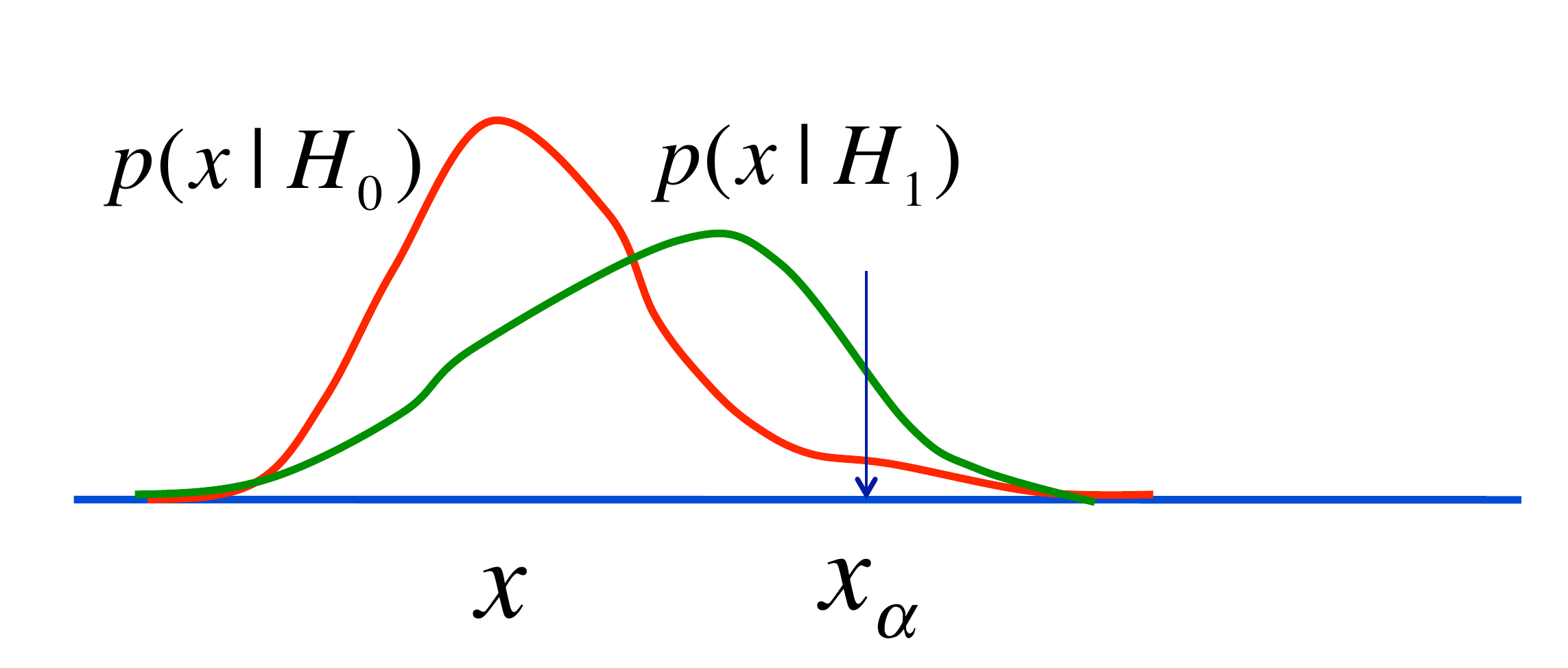}
	\caption{See Fig.~\ref{fig:neymantest1} for details. Unlike the case in Fig.~\ref{fig:neymantest1}, the  two hypotheses $H_0$ and $H_1$ are not that different. It is then
	not clear whether it makes practical sense to reject $H_0$ when $x > x_\alpha$ only
	to replace it with an hypothesis $H_1$
	that is not much better.}
	\label{fig:neymantest2}
\end{wrapfigure}

However, since
we have specified an alternative hypothesis there is more that can be said. Figure~\ref{fig:neymantest1} shows that we can also calculate
\begin{align}
	\beta & = P( x \leq x_\alpha | H_1),
\end{align}
which is the relative frequency with which we would reject the hypotheses of the form of $H_1$ if they are true.
These mistakes are called \textbf{Type II} errors. The quantity $1 - \beta$ is called the
\textbf{power} of the test and is the relative frequency with which we would accept 
$H_1$ if  true. Obviously, for a given $\alpha$ we want to maximize the power. Indeed, this
is  the basis of the Neyman-Pearson lemma (see for example Ref.~\cite{James}), which asserts that given two simple hypotheses --- that is, hypotheses in which all parameters have well-defined values --- the optimal statistic $t$ to use in the hypothesis test is the likelihood ratio
$t = p(x|H_1) / p(x | H_0)$. 

Maximizing the power seems sensible. Consider
Fig.~\ref{fig:neymantest2}. The significance of the test in this figure is the same as 
that in Fig.~\ref{fig:neymantest1}, so the Type I error rate is identical. However, the Type II error
rate is much greater in Fig.~\ref{fig:neymantest2} than in Fig.~\ref{fig:neymantest1}, that is, the power
of the test is considerably weaker in the former. In that case, there may be no compelling reason to reject the null since the alternative is not that much better. This insight was one source
of Neyman's disagreement with Fisher. Neyman objected to the possibility that one might reject a null hypothesis regardless
of whether it made sense to do so. He insisted that the task is always one of
deciding between competing hypotheses. Fisher's counter argument was that an alternative
hypothesis may not be available, but we may nonetheless wish to know whether the only
hypothesis that is available is worth keeping. As we shall see, the Bayesian approach
also requires an alternative, in agreement with
Neyman, but in a way that neither he nor Fisher agreed with!

We have assumed that the hypotheses $H_0$ and $H_1$ are simple, that is, fully specified. 
Unfortunately, most of the hypotheses that arise in realistic particle physics analyses are not of this kind. In the Higgs boson discovery analyses by ATLAS and CMS the probability models depend on many nuisance parameters for which only estimates are available. Consequently, neither the background-only nor the background $+$ signal hypotheses are fully specified.
Such hypotheses are called
\textbf{compound hypotheses}. 
  In order
to illustrate how hypothesis testing proceeds in this case,  we again turn again to the top discovery example.

\begin{quote}
\paragraph*{Example}

As we saw  in Sec.~\ref{sec:profile}, the standard way to handle nuisance
parameters in the frequentist approach is to replace them by their conditional MLEs and thereby reduce the likelihood 
function to the profile likelihood. In the top discovery example, we obtain a function $p_{PL}(D | s)$ that depends on the single
parameter,  $s$.  We now treat this function as if it were a likelihood and
appeal to both the Neyman-Pearson lemma, which suggests the use of likelihood ratios, and Wilks' theorem to motivate the  use of the 
function $t(x, s)$ given in Eq.~(\ref{eq:wilks}) to distinguish between two hypotheses:
the hypothesis $H_1$ in which $s = \hat{s} = N - B$ and the hypothesis $H_0$ in which $s \neq \hat{s}$, for example,
the background-only hypothesis $s = 0$. In the context of testing, $t(x, s)$ is called
a \textbf{test statistic}, which, unlike a statistic as we have defined it (see Sec.~\ref{sec:statistics}), usually depends on at least one unknown parameter.

In principle, the next step is the computationally arduous task of simulating the distribution
of the statistic $t(x, s)$. The task is arduous because \emph{a priori} the probability density 
$p(t| s, b)$ can depend on \emph{all} the parameters
that exist in the original likelihood. If this is  the case, then after all this effort  we seem to have achieved a pyrrhic victory! But, this is where Wilks' theorem saves the day, at least approximately. We can avoid the burden of simulating $t(x, s)$ because the latter is
approximately a $\chi^2$ variate.

Using $N = 17$ and $s = 0$, we find $\sqrt{t_0} = \sqrt{t(17, 0)} = 4.6$. According to the
results shown in
Fig.~(\ref{fig:toppl})(a),  $N = 17$ may 
can be considered ``a lot of data"; therefore, we may use $t_0$ to implement a hypothesis test by comparing $t_0$ with a fixed value
$t_\alpha$ corresponding to the significance level $\alpha$ of the test.

\end{quote}

\section{Lecture 3: the Bayesian approach}
In this lecture, we introduce the Bayesian approach to inference, again using the top quark
discovery data from D\O\ to illustrate the ideas. 

The Bayesian approach is merely applied
probability theory (see Section~\ref{sec:prob}).
A method is Bayesian if
\begin{itemize}
	\item it is based on the degree of belief interpretation of probability  and
	\item it uses Bayes theorem
	\begin{align}
		p(\theta, \omega | D) & = \frac{p(D|\theta, \omega) \, \pi(\theta, \omega)}{p(D)}, \\
		\textrm{where} \nonumber\\
				D 		& = \textrm{ observed data}, \nonumber \\
		\theta	& = \textrm{ parameters of interest}, \nonumber\\
		\omega	& = \textrm{ nuisance parameters}, \nonumber\\
		p(\theta, \omega| D) & = \textrm{posterior density}, \nonumber\\
		\pi(\theta, \omega) & = \textrm{prior density (or prior for short).} \nonumber
	\end{align}
\end{itemize}
for \emph{all} inferences. The result of a Bayesian inference is the posterior density
$p(\theta, \omega | D$ from which, if desired, various summaries can be extracted. The parameters can be discrete or continuous and nuisance parameters are eliminated by 
marginalization,
\begin{align}
	p(\theta | D) 	& = \int p(\theta, \omega | D ) \, d\omega, \\
				& \propto \int p(D | \theta, \omega) \, \pi(\theta, \omega) \, d\omega. \nonumber
\end{align}
The function $\pi(\theta, \omega)$, called the prior, encodes whatever information we have
about the parameters $\theta$ and $\omega$ independently of the data $D$. A key
feature of the Bayesian approach is recursion: the use of
the posterior density $p(\theta, \omega|D)$ or one, or more, of its marginals such as
$p(\theta| D)$ as the prior  in a subsequent analysis. 

These simple rules yield an extremely powerful and general inference model, a model
that was used, for example, in  the discovery of single top quark
production at the Tevatron~\cite{Abazov:2009ii, Aaltonen:2009jj}.

\subsection{Model selection}
Conceptually, hypothesis testing in the Bayesian approach (also called model selection)
proceeds exactly the same way as any other Bayesian calculation: we compute the 
posterior density,
\begin{align}
	p(\theta, \omega, H | D) 	& = \frac{p(D | \theta, \omega, H) \, \pi(\theta, \omega, H)} {p(D)},
\end{align}
and marginalize it with respect to all parameters except the ones that  label
the hypotheses or models, $H$, 
\begin{align}
	p(H | D ) & = \int p(\theta, \omega, H | D) \, d\theta \, d\omega.
	\label{eq:pHD}
\end{align}
Equation~(\ref{eq:pHD}) is
the probability of hypothesis $H$ given the observed data $D$.
In principle, the parameters $\omega$ could also depend on $H$. For example, suppose
that $H$ labels different parton distribution function (PDF) models, say CT10, MSTW, and
NNPDF, then $\omega$ would indeed depend on the PDF model and should be written
as $\omega_H$.

Like a Ph.D., it is usually more convenient to arrive at the probability $p(H|D)$ in stages.
\begin{enumerate}
	\item Factorize the prior in the most convenient form,
		\begin{align}
			\pi(\theta, \omega_H, H) & = \pi(\theta, \omega_H | H) \, \pi(H), \nonumber\\
				& = \pi(\theta |\omega_H, H) \, \pi(\omega_H | H) \, \pi(H),\\
				\textrm{or} \nonumber\\
				& =  \pi(\omega_H |\theta, H) \, \pi(\theta  | H) \, \pi(H).
		\end{align}
		Often, we can assume that the parameters of interest $\theta$ are independent,
		\emph{a priori}, of both the nuisance
		parameters $\omega_H$ and the model label $H$, in which case we can write,
		$\pi(\theta, \omega_H, H) = \pi(\theta) \, \pi(\omega_H|H) \, \pi(H)$.
		
	\item Then, for each hypothesis, $H$, compute the function
		\begin{align}
			p(D | H ) = \int p(D | \theta, \omega_H, H) \, \pi(\theta, \omega | H) \, d\theta \,
			d\omega.
		\end{align}
	\item Then, compute the probability of each hypothesis,
		\begin{align}
			p(H | D ) =\frac{p(D | H) \, \pi(H)} {\sum_H p(D | H) \, \pi(H)}.
		\end{align}	
\end{enumerate}
Clearly, in order to compute $p(H | D)$ it is necessary to specify the priors $\pi(\theta, \omega | H)$ and $\pi(H)$. With some effort, it is possible to arrive at an acceptable form for
$\pi(\theta, \omega | H)$, however, it is highly unlikely that consensus could ever be reached on the discrete prior
$\pi(H)$. At best, one may be able to adopt a convention.  For example, if by convention two hypotheses $H_0$ and $H_1$ are to be regarded as equally likely, \emph{a priori},
then it would make sense to assign $\pi(H_0) = \pi(H_1) = 0.5$.

One way to circumvent the specification of the prior $\pi(H)$ is to compare the probabilities,
		\begin{align}
			\frac{p(H_1 | D )}{p(H_0 | D)} =\left[ \frac{p(D | H_1)}{p(D | H_0} \right] \,
			 \frac{ \pi(H_1)} {\pi(H_0)}.
		\end{align}
and use only the term in brackets, called the global \textbf{Bayes factor}, $B_{10}$, as a way to
compare hypotheses. The Bayes factor specifies by how much the relative probabilities
of two hypotheses changes as a result of incorporating new data, $D$. The word global 
indicates that we have marginalized over all the parameters of the two models. The \emph{local}
Bayes factor, $B_{10}(\theta)$ is defined by
\begin{align}
	B_{10}(\theta) & = \frac{p(D| \theta, H_1)}{p(D| H_0)}, \\
	\textrm{where}, \nonumber\\
	p(D| \theta, H_1) & \equiv \int p(D | \theta, \omega_{H_1}, H_1) \, \pi(\omega_{H_1} | H_1) \, d\omega_{H_1},
\end{align}
are the \textbf{marginal} or integrated likelihoods in which we have assumed the \emph{a priori}
independence of $\theta$ and $\omega_{H_1}$. We have further assumed
that the marginal likelihood $H_0$ is independent of $\theta$, which is a very
common situation. For example, $\theta$ could be the expected signal count $s$,
while $\omega_{H_1} = \omega$ could be the expected background $b$. In this case, the
hypothesis $H_0$ is a special case of $H_1$, namely, it is the same as $H_1$ with $s = 0$. An hypothesis that is a special case of another 
 is said to be \textbf{nested} in the more general hypothesis. The example, discussed below, will
make this clearer. 

 There is a notational subtlety that may be missed:  because of the way we have
defined $p(D|\theta, H)$, we
need to multiply $p(D| \theta, H)$ by the prior $\pi(\theta)$ and then integrate with respect
to $\theta$ in order to calculate $p(D | H)$.

\subsubsection{Priors}
Constructing a prior for nuisance parameters is generally neither controversial (for most parameters) nor problematic. 
The Achilles heal of the Bayesian approach is the need to specify the prior $\pi(\theta)$,
for the parameters of interest,
at the start of the inference chain when we know almost nothing
about them. Careless specification of this prior can yield
results that are unreliable or even nonsensical. The mandatory requirement is that 
the posterior density be \textbf{proper}, that is integrate to unity. Ideally, the same should hold
for priors. A very extensive literature exists on the topic of prior specification
when the available information is extremely limited. However, a discussion of this
topic is beyond the scope of these lectures.

For model selection, we need to proceed with caution because
Bayes factor are sensitive to the choice of priors and therefore less robust than posterior densities. Suppose that the prior $\pi(\theta) = C f(\theta)$, where $C$ is a normalization
constant. The global Bayes factor for the two hypotheses $H_1$ and $H_0$ can be written as
\begin{align}
	B_{10} = C \frac{\int p(D | \theta, H_1) \, f(\theta) \, d\theta}{p(D | H_0)}.
\end{align}
Therefore, if the constant $C$ is ill defined, typically because $\int f(\theta) \, d\theta = \infty$,
the Bayes factor will likewise be ill defined. For this reason, it is generally recommended
that an improper prior not be used for parameters $\theta$ that occur only in one hypothesis, here $H_1$. However, for parameters that are common to all hypotheses, it is permissible to
use improper priors because the constants cancel in the Bayes factor.

The discussion so far has been somewhat abstract. The next section therefore works through an example of a possible Bayesian analysis of the D\O\ top discovery data.

\subsection{The top quark discovery: a Bayesian analysis}
In this section, we shall perform the following calculations as a way to illustrate a typical
Bayesian analysis,
\begin{enumerate}
	\item compute the posterior density $p(s | D)$,
	\item compute a 68\% credible interval $[l(D), u(D)]$, and
	\item compute the global Bayes factor $B_{10} = p(D | H_1) / p(D | H_0)$.
\end{enumerate}

\subsubsection*{Probability model}
The first step in any serious statistical analysis is to think deeply about  what has been done in
the physics analysis and construct a probability model. The full probability model is the joint
probability,
\begin{align*}
	p(x, s, b | I),
\end{align*}
which, as is true of \emph{all} probability models, is conditional on the information and assumptions,  $I$, that define the abstract space $\Omega$ (see Sec.~\ref{sec:prob}). 
In these lectures, we have
omitted the conditioning data $I$, but it should not
be forgotten that it is always present and may differ from one probability model to another.

The full probability model $p(x, s, b)$ can be factorized in several mathematically valid ways. However, we find it convenient to factorize the model in the following way,
\begin{align}
	p(x, s, b) = p(x | s, b) \, \pi(s, b),
\end{align}
where we have introduced the symbol $\pi$ in order to highlight the distinction we choose
to make between this
part of the model and the remainder. We shall compute the likelihood from $p(x|s, b)$
and view $\pi(s, b)$ as the prior for $s$ and $b$. We assume $p(x | s, b)$ to be
\begin{align}
p(x|s, b) = \textrm{Poisson}(x, s + b).
\label{eq:pxsb}
\end{align}
 The interpretation
of $p(x | s, b)$ is clear: it is the probability to observe $x$ events \emph{given} that the mean event count is $s + b$. 
What does $\pi(s, b)$ represent? This prior encodes what we \emph{know}, or \emph{assume}, about the mean background and signal independently
of the potential observations $x$. The prior $\pi(s, b)$ can be factored in two ways,
\begin{align}
	\pi(s ,  b) 	& = \pi(s | b ) \, \pi(b), \nonumber \\ 
			& = \pi(b | s ) \, \pi(s).
\end{align}
The factorizations remind us that the parameters
$s$ and $b$ may not be  independent. However, we shall assume that they are, at least at this stage of the analysis, in which case it is permissible to write,
\begin{align}
	\pi(s ,  b) 	& = \pi(s) \, \pi(b).
	\label{eq:prior1} 
\end{align}

\begin{wrapfigure}{R}{0.5\textwidth}
\centering\includegraphics[width=0.5\textwidth]{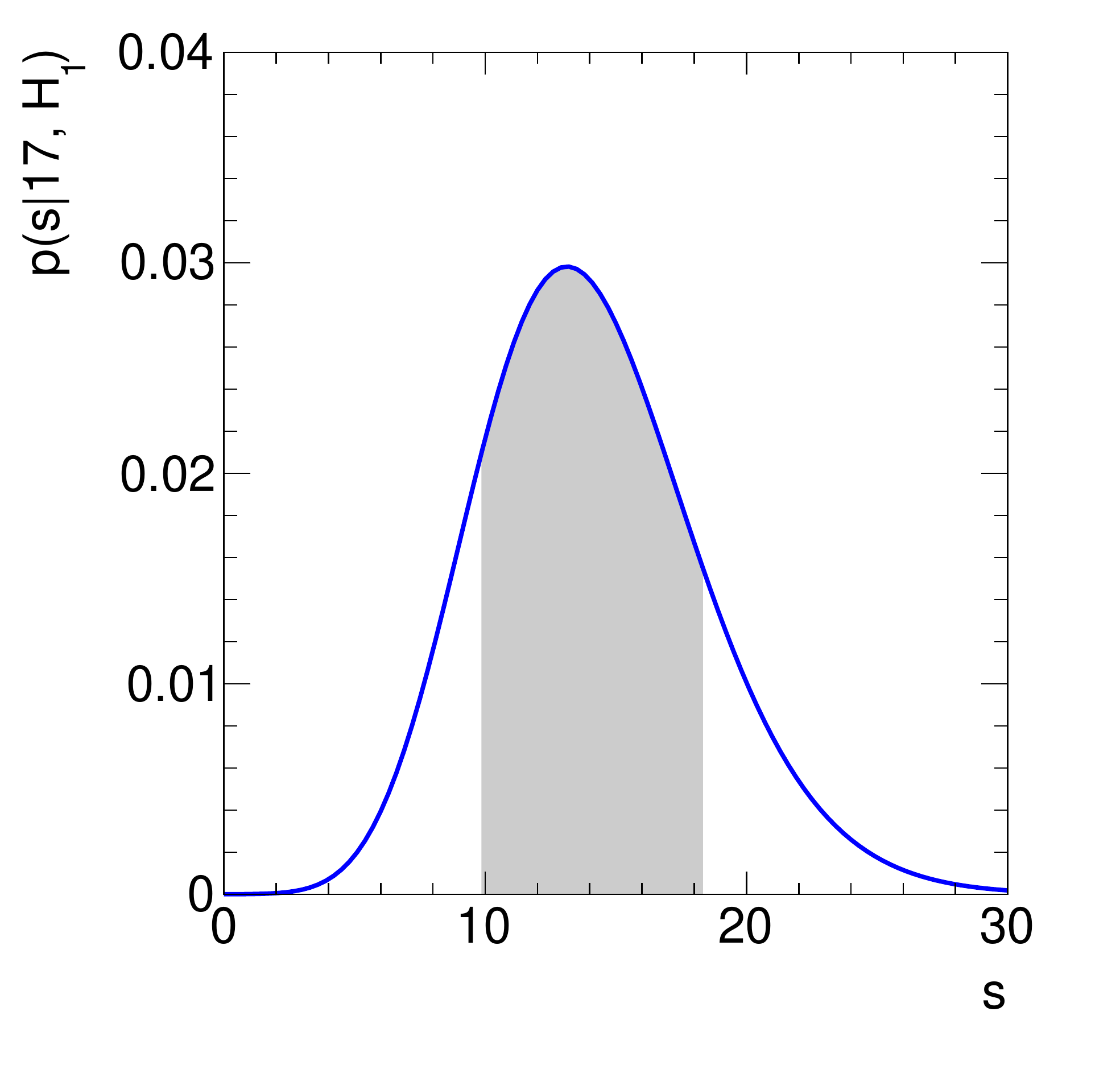}
\caption{Posterior density computed for D\O\ top quark discovery data. The shaded area
is the 68\% central credible interval.}
\label{fig:post}
\end{wrapfigure}

What do we
know about the background?
We know the count $Q$ in the control region and 
we have an estimate of the
control region to
signal region scale factor $k$. Since $Q$ is a count, a reasonable
model for the likelihood is 
\begin{align}
p(Q | k, b) = \textrm{Poisson}(Q, k b),
\label{eq:pQkb}
\end{align}
from which, together with a prior $\pi(k, b)$, we can compute the posterior density
\begin{align}
	p(b | Q, k) = p(Q | k, b) \, \pi(k, b) / p(Q, k).
	\label{eq:pbQk}
\end{align}
As usual, we factorize the prior, $\pi(k, b) = \pi(k|b) \pi_0(b) $,
where we have introduced the subscript $0$ to distinguish  $\pi_0(b)$  from the background prior
associated with Eq.~(\ref{eq:pxsb}). But, now we
need to construct $\pi(k|b)$ and $\pi_0(b)$ using whatever information we have at hand.
 
Clearly, $b \geq 0$. But, that miserable tidbit is all we know apart from the background likelihood, Eq.~(\ref{eq:pQkb})!
Today, after a century of
argument and discussion, the consensus amongst statisticians is that there is no
unique way to represent such vague information. However, 
well founded ways to construct such priors are available, see for example Ref.~\cite{Demortier:2010sn}
and references therein, but  for simplicity we take the prior $\pi_0(b) = 1$, that is, the \textbf{flat prior}. If the uncertainty in $k$ can be neglected, the (proper!) prior for $k$ is $\pi(k|b) = \delta(k - B/\delta B^2)$, which amounts to replacing $k$ in Eq.~(\ref{eq:pbQk}) by $B/\delta B^2$. This
yields,
\begin{align}
	p(b | Q, k) = \textrm{Gamma}(k b, 1, Q+1) = \frac{e^{-k b} (k b)^Q} {\Gamma(Q+1)},
\end{align}
for the posterior density of $b$, 
which can serve as the prior $\pi(b)$ associated with Eq.~(\ref{eq:pxsb}).

By construction, $p(x, s, b)$ is identical in form to the likelihood in Eq.~(\ref{eq:toplh}); we have 
simply availed ourselves of the freedom to factorize $p(x, s, b)$ as we wish and therefore to
reinterpret the factors. This freedom is useful because it makes it possible to keep the
likelihood simple while relegating the complexity to the prior. This may not seem, at first, to be terribly helpful; after all, we arrived at the same mathematical form as Eq.~(\ref{eq:toplh}). However, the complexity can be substantially mitigated by sampling
from the prior so that the model is represented by the relatively simple likelihood
and an ensemble of points that collectively represent the prior. The likelihood, as we have conceptualized the problem, is given by 
\begin{align}
	p(D| s, b) = \frac{e^{-(s+b)} (s + b)^D}{D!},
\end{align}
where $D = 17$ events. 

The final ingredient is the prior $\pi(s)$. At this stage, all we know is that $s \geq 0$ and, again,
there is no unique way to specify $\pi(s)$, though as noted there are well founded methods to
construct it. We shall assume either the improper prior $\pi(s) = 1$ or the proper prior $\pi(s) = \delta(s - 14)$.

\subsubsection*{Marginal likelihood}
We have done the hard part: building the full probability model. Hereafter, the rest of the Bayesian
analysis is mere computation.

It is convenient  to eliminate the nuisance parameter $b$,
\begin{align}
	p(D | s, H_1) 	& = \int_0^\infty p(D | s, b ) \, \pi(b ) d(k b),\nonumber\\
				& = \frac{1}{Q} (1- x)^2 \sum_{r=0}^N \textrm{Beta}(x, r+1, Q) \, 
				\textrm{Poisson}(N - r| s ),\\
	\textrm{where }	x & = 1/(1+k), \nonumber\\ \nonumber\\
				& \framebox{\textbf{Exercise 10:} Show this} \nonumber
\end{align}
and thereby arrive at the marginal likelihood $p(D | s, H_1)$.

\subsubsection*{Posterior density}
Given the marginal likelihood $p(D | s, H_1)$ and a prior $\pi(s)$ we can compute the posterior density,
\begin{align}
	p(s | D, H_1) 	& =  p(D | s, H_1) \, \pi(s) / p(D | H_1), \\
	\textrm{where,} \nonumber \\
				p(D  | H_1) & =  \int_0^\infty p(D | s, H_1) \, \pi(s) \, ds. \nonumber
\end{align}
Assuming a flat prior for the signal, $\pi(s) = 1$,we  
find
\begin{align}
	p(s | D, H_1) 	& =  \frac{\sum_{r=0}^N \textrm{Beta}(x, r + 1, Q) \, \textrm{Poisson}( N - r| s)}
	{\sum_{r=0}^N \textrm{Beta}(x, r + 1, Q)}, \\ \nonumber\\
		& \framebox{
			\parbox{0.6\textwidth}{\textbf{Exercise 11:} \textrm{Derive an expression for}	
				$p(s | D, H_1)$	
				assuming $\pi(s) = $ Gamma$(q s, 1, M + 1)$ where $q$ and $M$ are constants}
				}
			\nonumber
\end{align}
from which we can compute the central \textbf{credible interval} $[9.9 , 18.4]$ for $s$ at
68\% C.L., which is shown in Fig.~\ref{fig:post}. The statement $s \in [9.9, 18.4]$ at 68\% C.L.
means there is a 68\% probability that $s$ lies in $[9.9, 18.4]$. Unlike the frequentist
statement, this statement is about this particular interval and the 68\% is a degree of
belief, not a relative frequency. That being said, the best Bayesian methods tend to
produce credible intervals that also approximate confidence intervals.

\subsubsection{Bayes factor}
As noted, the number $p(D | H_1)$ can be used to perform a hypothesis test. But, as argued
above, we need to use a proper prior for the signal, that is, a prior that integrates to one.
The simplest such prior is a $\delta$-function, e.g., $\pi(s) = \delta(s - 14)$. Using this prior,
we find
\begin{align*}
	p(D | H_1) = p(D | 14, H_1)  = 9.28 \times 10^{-2}.
\end{align*}
Since the background-only hypothesis $H_0$ is nested in $H_1$, and defined by $s = 0$, the number $p(D | H_0)$  is given by  $p(D|0, H_1)$, which yields
\begin{align*}
	p(D | H_0) = p(D | 0, H_1)  = 3.86 \times 10^{-6}.
\end{align*}
We conclude that the hypothesis $s = 14$  is favored over $s = 0$ by a Bayes factor of 24,000. In order to avoid large numbers, the Bayes factor can be mapped into a (signed) measure akin
to the frequentist ``$n$-sigma"~\cite{Sezen},
\begin{align}
	Z = \textrm{sign}(\ln B_{10}) \sqrt{2 |\ln B_{10}|}, 
\end{align}
which gives $Z = 4.5$. Negative values of $Z$ correspond to hypotheses that are excluded.

\section*{Summary}
We have given an overview of the main ideas of statistical inference in a form
directly applicable to statistical analysis in particle physics. Two widely used approaches
were covered, frequentist and Bayesian.  Statistics is not
physics.  While Nature is the ultimate arbiter of which physics ideas are ``correct", the ultimate arbiter of statistical ideas is intellectual taste. Therefore, we hope you take to heart the following
advice.

\begin{quote}
	``Have the courage to you use your own understanding"
	\smallskip
	
	Immanuel Kant
\end{quote}

\section*{Acknowledgement}
I thank Nick Ellis, Martijn Mulders, Helene Haller, and their counterparts from JINR, for
organizing and hosting an excellent school, and the students for their
energetic participation. These lectures were supported in part by
US Department of Energy grant DE-FG02-97ER41022.

\end{document}